# FLOW DESCRIPTORS OF HUMAN MOBILITY NETWORKS

David Pastor-Escuredo[1,2,*,] Enrique Frias-Martinez[3]

[1]LifeD Lab, Madrid, Spain
[2]Center Innovation and Technology for Development, Technical University Madrid, Spain.
[3]Telefónica Research, Madrid, Spain
[*]email: david@lifedlab.org

**ABSTRACT**

The study of human mobility is key for a variety of problems like traffic forecasting, migration flows of virus spreading. The recent explosion of geolocated datasets has contributed to better model those problems. In this context, mobile phone datasets enable the timely and fine-grained study human mobility, allowing the description of mobility at different resolutions and with different spatial, temporal and social granularity. In this paper we propose a systematic analysis to characterize mobility network flows and topology and assess their impact into individual traces. Discrete flow-based descriptors are used to classify and understand human mobility patterns at multiple scales. This framework is suitable to assess urban planning, optimize transportation, measure the impact of external events and conditions, monitor internal dynamics and profile users according to their movement patterns.

**INTRODUCTION**

Since over more than a decade, mobile phone data has enabled for the analysis of individual and collective human mobility (Gonzalez et al., 2008) (Candia et al., 2008). Several studies have focused on the models to predict mobility from big data sources and theoretical and practical limits of predictability (Song et al., 2010) (Simini et al., 2012) (Lu et al., 2013). The analysis of mobility has permitted novel studies and applications in the intersection with social science (Blondel et al., 2015) (Naboulsi et al., 2015).

A relevant focus of research and application has been the humanitarian sector. Mobile phone data has shown its potential to unravel mobility patterns suitable to understand and prevent from the spread diseases outbreaks (Bengtsson et al., 2011) (Wesolowski et al., 2012) (Wesolowski et al., 2015) (Heesterbeek et al., 2015) (Bengtsson et al., 2015) and support humanitarian action during and after natural disasters (Bagrow et al., 2011) (Lu et al., 2012) (Pastor-Escuredo et al., 2014) (Ghurye et al., 2016) (Wilson et al., 2016) (Pastor-Escuredo et al., 2018).

Mobile phone data has also served for development as proxy for mapping poverty and estimating socio-economic profile of population groups (Eagle et al., 2010) (Soto et al., 2011) (Blumenstock et al., 2015) (Blumenstock, 2016) (Pappalardo et al., 2016) (Pokhriyal and Jacques, 2017) (Steele et al., 2017), understanding food security and vulnerability to climate change (Decuyper et al., 2014) (Lu et al., 2016) (Zufiria et al.,

2018) or helping to optimize infrastructure development (Çolak et al., 2015). Cities have been also studied from the lens of mobile phone data, not only to understand transportation but to investigate social dynamics (Barlacchi et al., 2015) (De Nadai et al., 2016) (De Nadai, 2019).

Typically, mobility characterization is expressed in terms of origin-destination matrices that do not convey information about individual trajectories through time (Cascetta et al., 1993) (Ma et al., 2013) (Iqbal et al., 2014) (Alexander et al., 2015). Another approach is to build a network of displacements between nodes (locations) aggregating mobility (Morales, 2015). Understanding the propagation in networks has been addressed through routing algorithms that assess how the information travels within the network (Watts and Strogatz, 1998) (Watts et al., 2002) (Dodds et al., 2003) (Liben-Nowell et al., 2005) (Herrera-Yagüe et al., 2015). However, routing algorithms are a statistical tool that relies of averages of patterns of the real paths through the network. The analysis of individual trajectories allows long-term observation suitable to profile users and assess their socio-economic profile and potential vulnerability (Zufiria et al., 2018), but in order to understand mobility factors it is necessary to relate the collective mobility represented by the network with the movements of individuals. Field theory has been recently applied to study human mobility (Mazzoli et al., 2019). However, human mobility is not only determined by a Euclidean distance space, but largely influenced by a "flow space" between the nodes of a network that depends on the traction of the nodes.

In this work, we propose a systematic framework to quantify the influence of large-scale mobility network characteristics into human trajectories. For this purpose, we compute flow-based and distance-based network topology descriptors characterizing the nodes and edges of the mobility network built from phone traces. We then express trajectories in terms of the descriptors. Sequences of descriptor gradients based on relative displacements and locations changes referred to the user`s home location allowed a reconstruction of dynamics integrating individual and collective patterns. Thus, the framework permits constructing descriptors at multiple time scales. We applied the framework to the analysis of mobility in the cities of Bogota and Medellin in Colombia. This framework is suitable to assess urban planning, optimize transportation, measure the impact of external events and conditions, monitor internal dynamics and profile users according to their movement patterns.

**METHODS**

**Building trajectory and identifying home location**

A dataset containing aggregated and encrypted mobility data for a period of six months was used to model trajectories. Trajectories were built by gathering all information belonging to the same user along time. This representation allowed having both displacements and long-term observation of users' trajectories.

Each user was assigned a home location. This location was computed, as proposed by the literature, as the most visited location from 8 pm for each user during a chosen reference week. If no CDRs are available for a user in this time interval, the most visited location for the whole week was chosen as alternative home location.

**Flow-based and distance-based network descriptors**

*Building mobility network*

Trajectories were used to build a network of users' displacements with a daily resolution. The network was modelled by both a directed graph and an undirected graph with a property "flow" to account for the people traveling between nodes (antenna locations). For the undirected graph, the flow along a network edge was incremented any time a user travelled between two antennas. For the directed graph, the flow was incremented in the direction of the displacement of the user. This process was repeated for all displacements during a day. The "inverse flow" was computed to have a metric of the resistance between nodes in terms of flow, necessary to compute current flow centrality metrics as described later on. As the mobility network is a graph that can be projected into the geographical space through the coordinates of the antennas represented by the nodes, physical variables (space and time) can be added to the description of the edges of the network. The descriptors added were the distance of the edge and average time consumed by a user to travel along the two nodes connected by the edge. Thus, the edges of the network were described with the following descriptors:

| **Flow** | The number of users traveling between the two nodes connected by the edge during a day. In the directed graph, two different directed flow-edges are distinguished according to the direction of the movements. |
|---|---|
| **Inverse flow** | The inverse of the flow for minimization of the flow through the network necessary for computation of current flow centrality. In the directed graph, two different directed inverse flow-edges are distinguished according to the direction of the movements. |
| **Distance** | The distance between the two nodes that correspond to the physical distance between the two antennas represented by the nodes. |
| **Average time** | The average time for a user to travel between the two nodes that correspond to the average travel between the two antennas for all the flow. The average was computed for all travels within a day. In the directed graph representation, the average time is differentiated according to the direction of the movements. |

*Table 1: Mobility network edge descriptors*

Of note, the average time of the travels along a graph edge is the most unstable parameter as it depends on the frequency of calls of the users that impose a subsampling in the mobility.

Networks were filtered to account only for the displacements within the region of interest defined by the antennas belonging to the cities considered: Bogotá and Medellín. Further processing was applied to obtain the largest connected network for the undirected graph and the largest weakly connected network for the directed graph. This processing was necessary to apply current flow centrality algorithms.

*Flow-based and distance-based centrality descriptors*

The topology of the network provides information of the large scale organization of the mobility. Trajectories within the networks can be classified as network walks, meaning that within a trajectory, network nodes and edges can be visited and traversed multiple times (Borgatti, 2005). Of note, trajectories projected into the geographical space may often describe shortest paths in space, but that does not imply network geodesics being these the shortest paths in the network characterized by the its flow property. Furthermore, as the mobile phone data provides a subsampled representation of the real mobility, several travels through edges of the network walks may not be present. This is an important characterization of mobility flows represented as a network.

For each node, centrality metrics (Freeman, 1977) (Bonacich, 1987) (Stephenson and Zelen, 1989) (Brandes, 2001) (Brandes and Fleischer, 2005) (Borgatti and Everett, 2006) (Newman, 2018) were computed. Based on volume of the flow in the network, degree centrality (1-step paths) and eigenvalue centrality (infinite paths) were computed. Shortest-path based centrality metrics were computed using the inverse flow property as the metric for characterizing paths: closeness centrality and betweenness centrality. Closeness and betweenness were computed also using the physical distance as the metric for shortest-paths. Finally, current flow centrality closeness and betweenness were computed using the inverse flow as the resistance property in network edges. Current flow centrality differs from shortest-path because it considers that the flow can split across several edges of a node as the electric current. To compute current flow metrics (Brandes and Fleischer, 2005), the largest connected undirected graph for the region of interest was chosen as base graph. Current flow centrality was only computed for the undirected graph. The following table summarize the descriptors attached to the network nodes to describe the topology and dynamics of the network flow.

| **In/out degree** | Number of directed edges arriving to/departing from a node in the directed graph |
|---|---|
| **Degree Centrality** | Radial and volume-based centrality computed from 1-length walks (normalized degree) based on the flow property. This centrality was computed for both directions of the directed graph. |
| **Eigenvalue Centrality** | Radial and volume-based centrality computed from infinite length walks. This centrality was computed for both directions of the directed graph. |
| **Closeness Centrality (flow)** | Radial and length-based centrality that considers the length of the shortest past of all nodes to the target node based on the flow property. This centrality was computed for both directions of the directed graph. |
| **Betweenness Centrality (flow)** | Medial and volume-based centrality that considers the number of shortest paths passing by a target node based on the flow property. This centrality was computed for both directions of the directed graph. |
| **Closeness Centrality (distance)** | Radial and length-based centrality that considers the length of the shortest past of all nodes to the target node based on the physical distance property. This centrality was computed for the undirected graph. |
| **Betweenness Centrality (distance)** | Medial and volume-based centrality that considers the number of shortest paths passing by a target node based on the physical distance property. This centrality was computed for the undirected graph. |
| **Current flow Closeness Centrality** | Radial and length-based centrality based on current flow model using the inverse flow property. This centrality was computed for the largest connected undirected subgraph. |
| **Current flow Betweenness Centrality** | Medial and volume-based centrality based on current flow model using the inverse flow property. This centrality was computed for the largest connected undirected subgraph. |

*Table 2: Network topology descriptors attached to the graph nodes*

The edge descriptors were used to obtain statistical descriptors at the node level:

| | |
|---|---|
| **Total in/out flow** | Number of users traveling from one node to one of all the neighbour nodes in the network (and in the reverse direction) |
| **Average in/out flow** | Average number of users traveling from one node to one of all the neighbour nodes in the network (and in the reverse direction) |
| **Average in/out distance** | Average of the distance of the travels between one node towards the neighbours in the network (and in the reverse direction) |
| **Average in/out time** | Average time of travel between one node and all its neighbours in the network (and in the reverse direction) |

*Table 3: Aggregated edge-based descriptors attached to the graph nodes*

**Eulerian and Lagrangian vectorizations of descriptors**

We used the users' trajectories to build Eulerian and Lagrangian vectors of flow-based and distance-based descriptors of the network. Trajectories are pathlines within the network flows. Thus, we generated time-evolving dynamic representation of the mobility network and the level of individuals. Three types of vectors were defined:

*Node-based vector*

A sequence of descriptors derived from the nodes visited by each user. This is a Eulerian (based on a spatial reference) representation of the dynamics of each user.

*Home-referenced vector*

Lagrangian representation of dynamics are based of the evolution regarding a reference state of flows and materials. We posit that this reference state in human mobility is the geospatial mapping of user's home location. Thus, we vectorize the gradient between the descriptors at each current position compared to the descriptors of the home location.

The expected daily evolution of this vectorization starts in the reference state and evolves to return to the reference state. Each value of the vector is computed in a different temporal interval possible thanks to the Lagrangian (trajectory-based) representation of the mobility.

*Displacement Gradient vector*

A sequence of the relative differences in the descriptors along the displacements of the users (the difference between the value of two nodes connected in the displacement). This is a quasi-Lagrangian representation as each vector has a different temporal scale (determined by the sampling of each user trajectory) and it is possible to estimate them thanks to the trajectory representation. Thus, this representation is suitable to make a multi-scale temporal analysis by connecting sampled user displacements to obtain a net displacement in specific time intervals.

**User's trajectory descriptors and homophily analysis**

The vectorization allows obtaining descriptors of the users' trajectories through time series analysis. These descriptors provide characteristic footprints of individual mobility integrating information of the network topology and dynamics. Simple descriptors for each user were computed as the statistics (mean, median, maximum, minimum and standard deviation) of the time series for each descriptor implying a scalarization. Thus, we reduced the dimensionality of representation of the mobility flows in terms of several network descriptors and their statistics. This approach also facilitated further processing of the different length vectors of each user that depended on the number of CDRs registered for each of them.

The scalarization allowed the classification of users into clusters and assess the variability within a specific location by gathering users according to their home location for homophily analysis.

**Visualization**

Visualization has been performed using CARTO. Maps based on the geolocation of the antennas have been created to display heatmaps for two types of representation:

- Descriptor values at a given node location.

- Homophily and statistics of the descriptors of a population based on a specific location

- Population counts of a segmented population group according to a given descriptor or set of descriptors.

**RESULTS**

**Flow-based centrality descriptors**

We propose a comprehensive framework of flow-based descriptors to characterize mobility. Human mobility is largely influenced by the social dimensions expressed in terms of flows besides physical dimensions (space and time) (Zufiria et al., 2018). The flows are determined by the directional movements between nodes of the network. By computing centrality metrics based on the *flow* and *inverse flow* properties of the network edges we can assess, at the level of each antenna, the influence of large scale topology and the whole mobility organization. We computed a reference network for each region (Bogotá and Medellín) as the average network (Supplementary Fig. 1-4) of the working days of a refence week (January 20$^{th}$ 2014). The reference networks were filtered to have edges with *weight* > 1.

The descriptors proposed (Methods) are non-redundant metrics of human mobility that can be further analysed at the level of the users by connecting the set of antennas that conform the individual trajectories. Figure 1 shows the correlation between pairs of descriptors as well as the histogram of each descriptor (main diagonal) for the reference network in Bogotá filtered on edges with *flow* > 1. The descriptors for the unfiltered reference network are shown in Supplementary Fig. 5. The results for Medellín are shown in Supplementary Fig. 6. It can be observed that only in/out values of the same descriptor display a linear correlation. Thus the information of the framework of descriptors is not redundant and each descriptor conveys valuable information to understand factors of mobility at scale. The histogram of the descriptors shows either an exponential decay with large concentrations near the zero and approximately gaussian distributions. These distributions are common in social parameters for large populations.

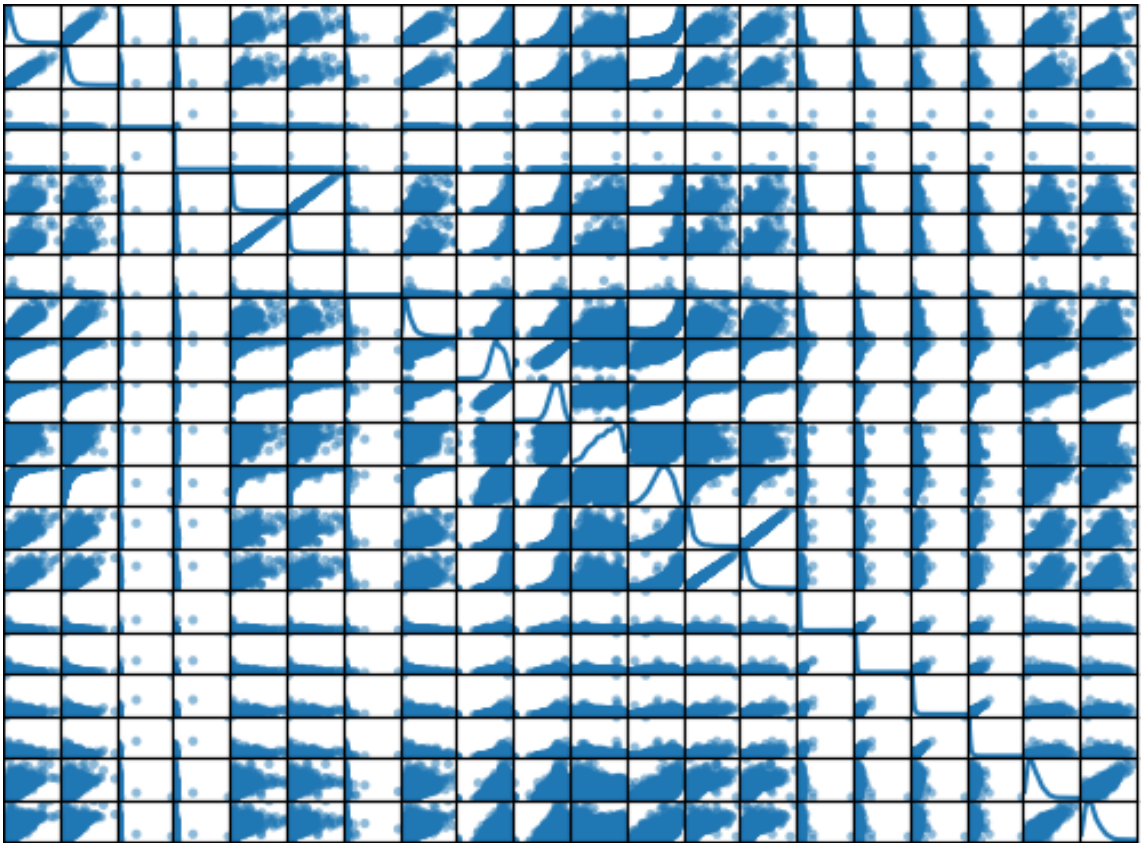

*Figure 1: Exploration of descriptors histogram and pair-wise correlation at the antenna level in Bogotá. The twenty descriptors computed are: 'in_degree', 'out_degree', 'in_eigenvalue', 'out_eigenvalue', 'in_betweenness', 'out_betweenness', 'dis_betweenness', 'cfbetweenness', 'in_closeness', 'out_closeness', 'dis_closeness', 'cfcloseness', 'in_flow', 'out_flow', 'in_ave_flow', 'out_ave_flow', 'in_std_flow', 'out_std_flow', 'in_ave_distance', 'out_ave_distance'.*

**Descriptor maps**

The flow-based and distance-based descriptors computed at the network nodes of the filtered reference network, including the connectivity filtering (Methods), were projected into a map to show the geographical distribution of the network properties (Fig 2). The maps of in flow, in degree, current flow betweenness an current flow closeness show a highly heterogeneous distribution within the city of Bogotá.

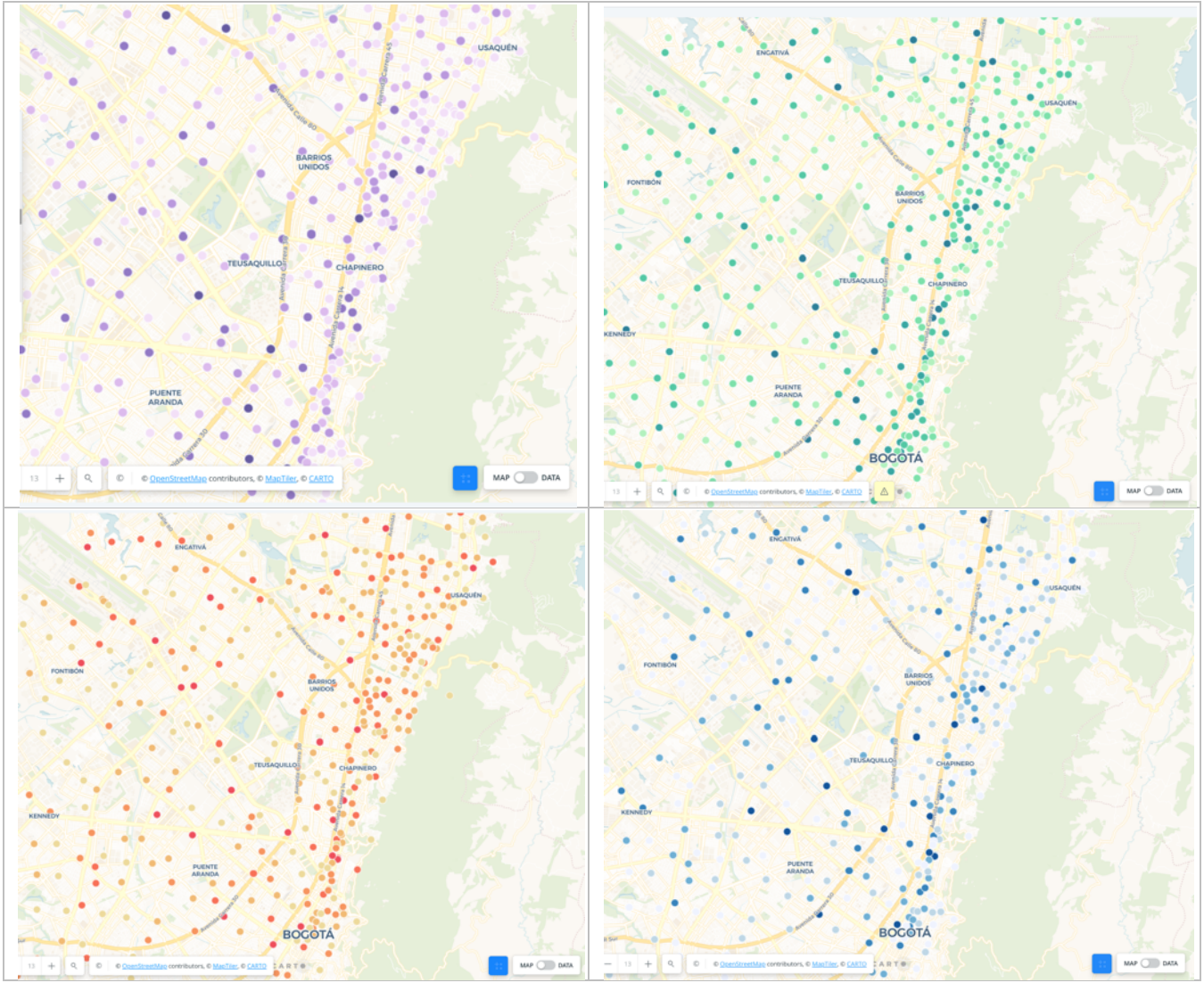

Figure 2: Top left: in_flow. Top right: in_degree. Bottom left: cfbetweenness. Bottom right: cfcloseness. Maps are from the city of Bogotá

The spatial heterogeneity of the descriptor and the lack of clear clusters within the city imply that there are fine grained factors that affect the topology of the human mobility network, for instance, the transportation system. Results show that the centrality of the nodes is distributed in different neighbourhoods and that centrality is a meaningful tool to investigate factors of mobility within areas and across areas of a city.

To minimize the effect of the spatial sampling due to the location of the antennas, the descriptors can be aggregated into larger scale grids. This visualization allows observing patterns and gradients at the city level (Fig. 3). Interestingly high values of cfcloseness are distributed in the central area of the city, but not as expected from distance-based closeness that concentrates in the city centre (Fig. 3). Further analysis of the flow-based indicators require understanding their temporal evolution and the variability and homophily within population groups that belong to each location.

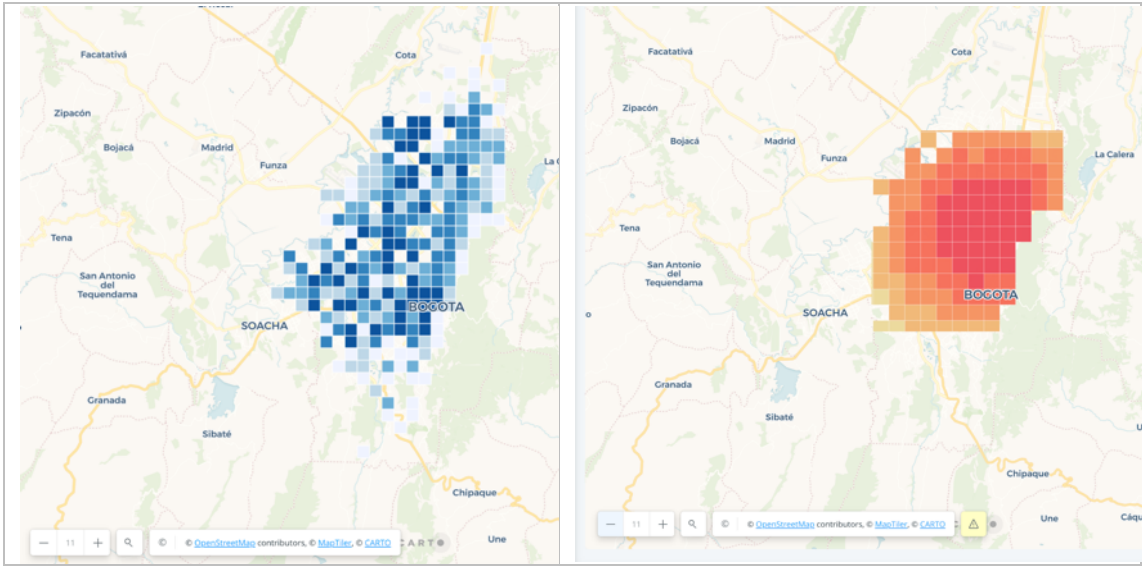

*Figure 3: Left: Aggregated cfcloseness. Right: Aggregated dis_closeness. Maps are displayed for the city of Bogotá*

**Vectorization of descriptors along user trajectories**

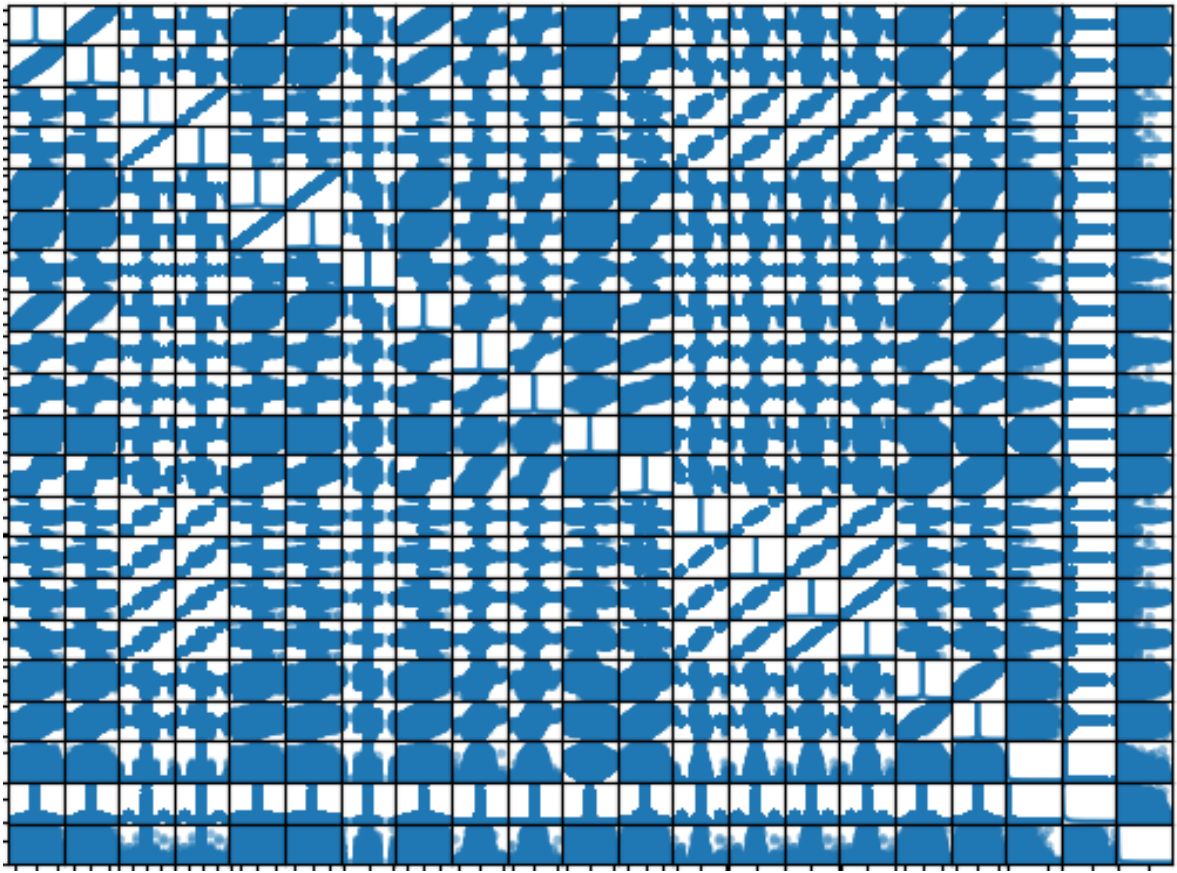

*Figure 4: Matrix of correlation and histograms for the gradient on network descriptors along user displacements. Descriptors in oder are: 'd_in_degree', 'd_out_degree',*

*'d_in_eigenvalue', 'd_out_eigenvalue', 'd_in_betweenness', 'd_out_betweenness', 'd_dis_betweenness', 'd_cfbetweenness', 'd_in_closeness', 'd_out_closeness', 'd_dis_closeness', 'd_cfcloseness0, 'd_in_ave_flow', 'd_out_ave_flow', 'd_in_std_flow' ,'d_out_std_flow', 'd_in_ave_distance', 'd_out_ave_distance', 'distance', 'flow', 'd_time'.*

By reconstructing users' trajectories it is possible to build temporal gradients of the descriptors based through the displacements between nodes. This allows estimating the potential correlation of descriptors to characterize why a user moves from one node to the other. Figure 4 shows the matrix of correlations and histograms for the gradient of descriptors. Of note, there is no correlation between the descriptors and the displacement characteristics (distance and trip time) or the edge characteristic flow. As in the description of the node properties, in and out measurements of the descriptors are highly correlated, indicating a symmetric behaviour of nodes. Also, eigenvalue centrality is very correlated with the difference in the average flow of the nodes in the direction of the displacement.

**Trajectory descriptors**

Further characterization of the mobility was made by computing statistics of the trajectory vector. The trajectory vector comprises the sequence of descriptor gradients of the displacements of the trajectory. Trajectories were computed through a week in January 2014 (20$^{th}$ January) and the descriptors were derived from the reference network computed for that same week. Trajectories were filtered to remove null vectors resulting from two consecutive identical locations in the trajectory of the users. Of note, this vectorization is based on single displacement reconstructions, this means the vector between two consecutive CDRs for a given user. Multi-scale reconstruction is possible by concatenating different displacements for the same user to modify the scale of the movement. It is also possible to make a temporal structural analysis by computing displacements for given temporal windows.

Figure 5 shows the distribution of the mean of several descriptor gradients along individual trajectories for Bogotá and Medellín. The distribution for both cities follow similar patterns potentially revealing invariant behaviours in people mobility. The distribution of the descriptors in each user follow a Gaussian-like structure centered in zero. This indicates that, for the most part of the people, most part of the trips are made between antennas with similar network properties. The tails represent users that have more distinctive mobility patterns so it is suitable to cluster different population groups.

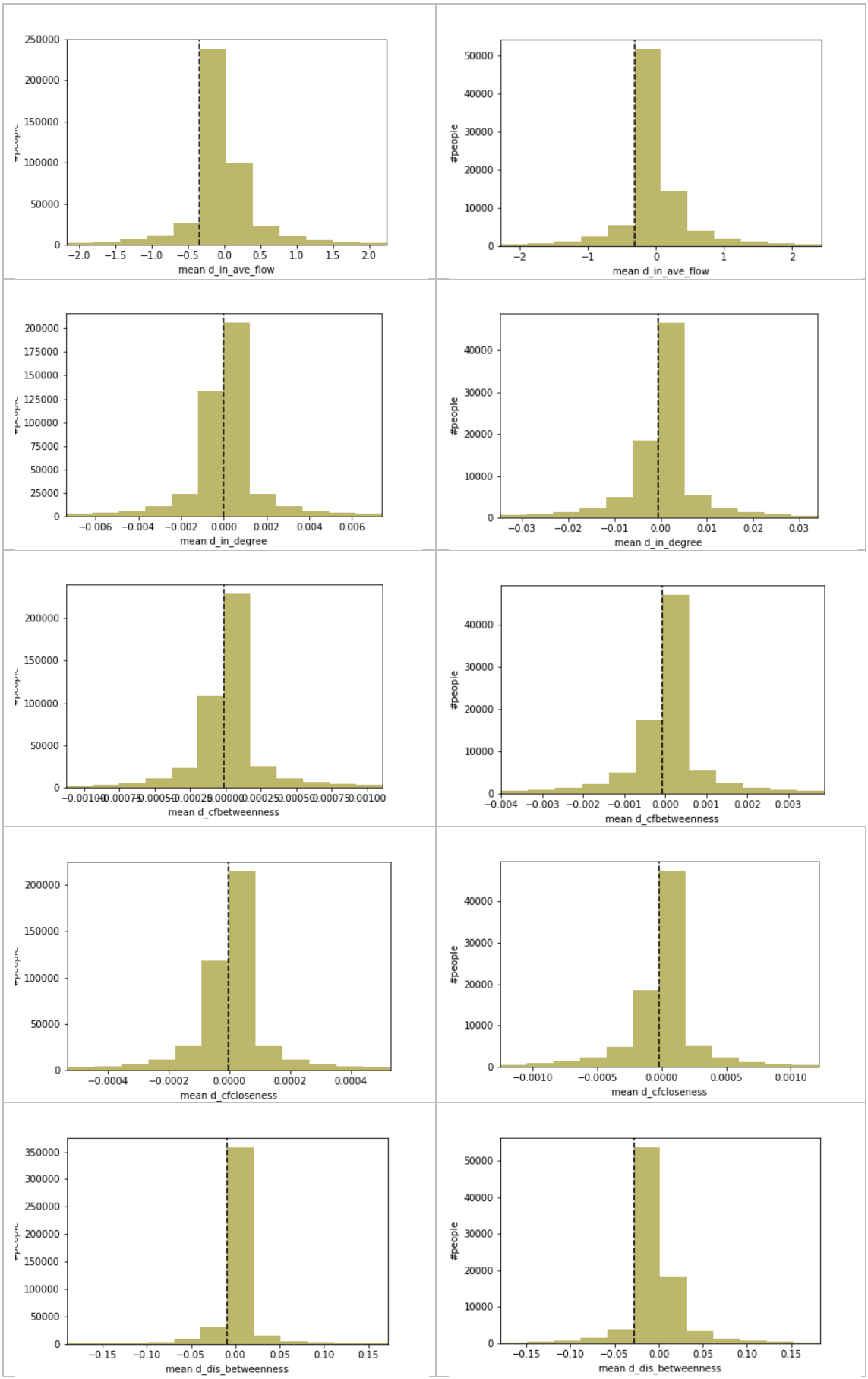

*Figure 5: Histograms of average of descriptors along users' trajectories. Left: Bogotá. Right: Medellín. First row: d_in_ave_flow. Second row: d_in_degree. Third row: d_cfbetweenness. Fourth row: d_cfcloseness. Fifth row: d_dis_closeness.*

**Spatial distribution of trajectory descriptors**

The characteristics of users' trajectories can be projected onto the geographical map by assigning each trajectory a home location (Methods). Thus, we can get statistics of the variability and homophily of users' behaviour within a specific geographical locations. Figure 6 shows that the average centrality along a trajectory is heterogeneously distributed as well. However, the variability represented by the standard deviation of the average profile of each trajectory is more localized in a specific area. This metric is potentially a very good proxy of social variability.

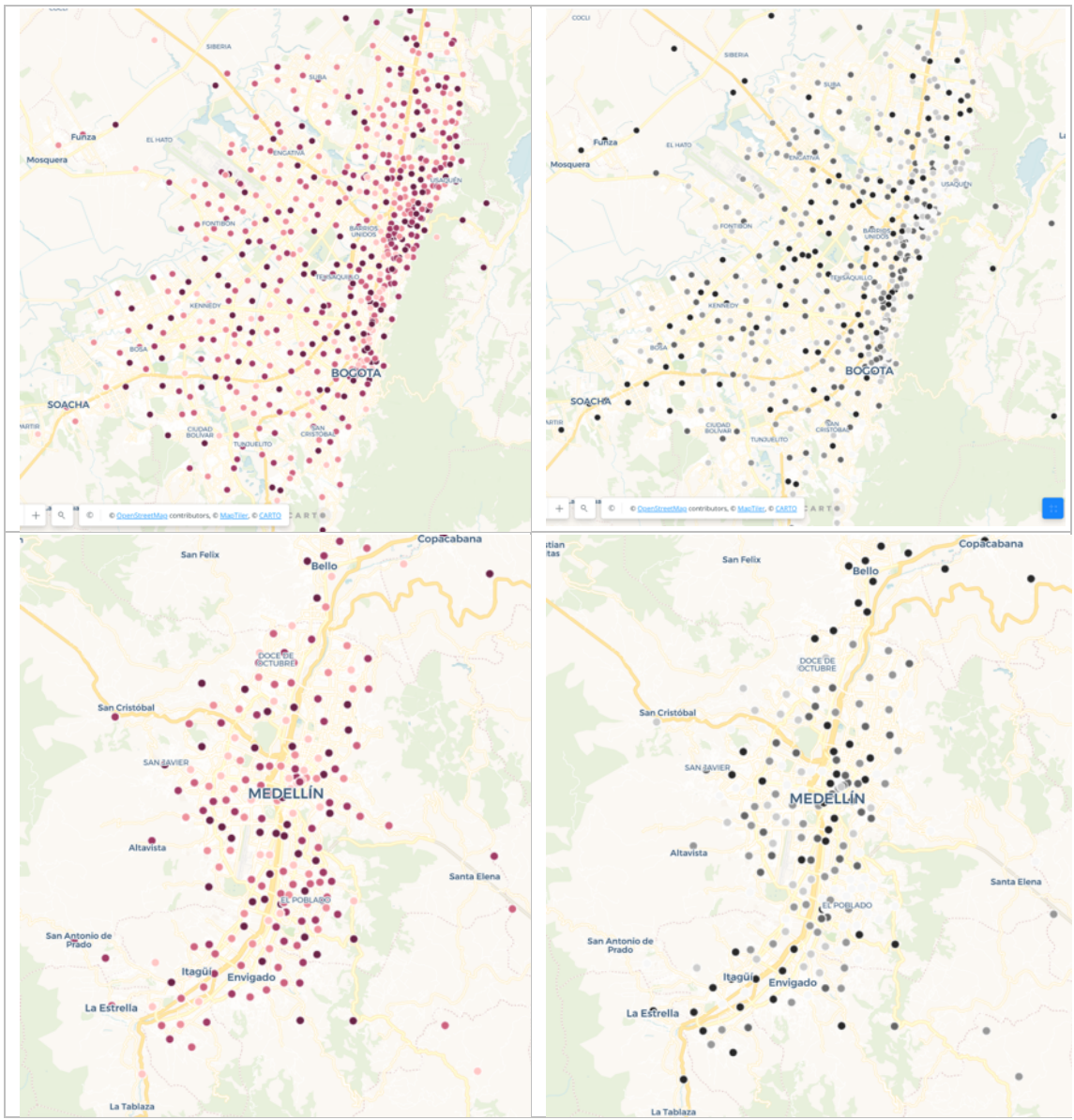

*Figure 6: Top: Bogotá. Bottom: Medellín. Left: average of in_degree mean along users trajectories: Right: std of in_degree mean along users trajectories*

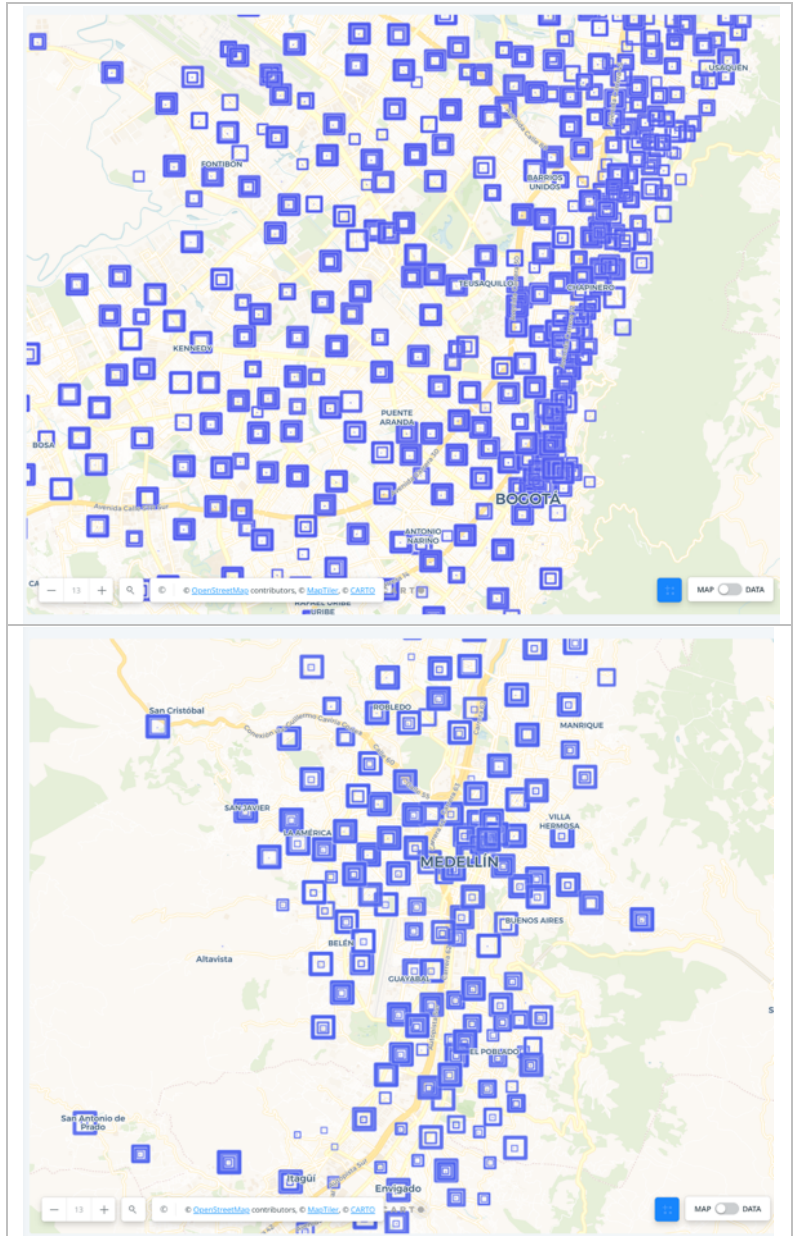

*Figure 7: Population count of users whose characteristic (average) d_in_degree are in the positive tail of the histogram distribution*

The distribution of Fig. 5 can be used to segment different population groups according to their characteristic descriptor. These population groups can be projected as well onto the geographical map after the assignment of home location for each user (Methods). Figure 7 shows the count of the population in the positive tail of the *d_in_degree* distribution (Fig. 5) in Bogotá and Medellín. As before, the distribution of the population belonging to this group is highly heterogeneous, but the data can be useful to characterize specific locations and track the target population along their trajectories (Pastor-Escuredo et al., 2015). In this sense, it is interesting to analyse if the home location and its properties influences the type of trajectory of the individuals. Further homophily could be approached by combining different descriptors into a feature vector for each user.

**Dynamics of user trajectories referred to the home location**

The vectorization of the descriptors along users' trajectories was made using the home location as initial point and the current position as the destination of the relative vector. Thus, instead of relative displacements we computed a sequence of vectors referred to the home location of each user. This (Lagrangian) representation is suitable to understand how home location affects to the mobility of individuals.

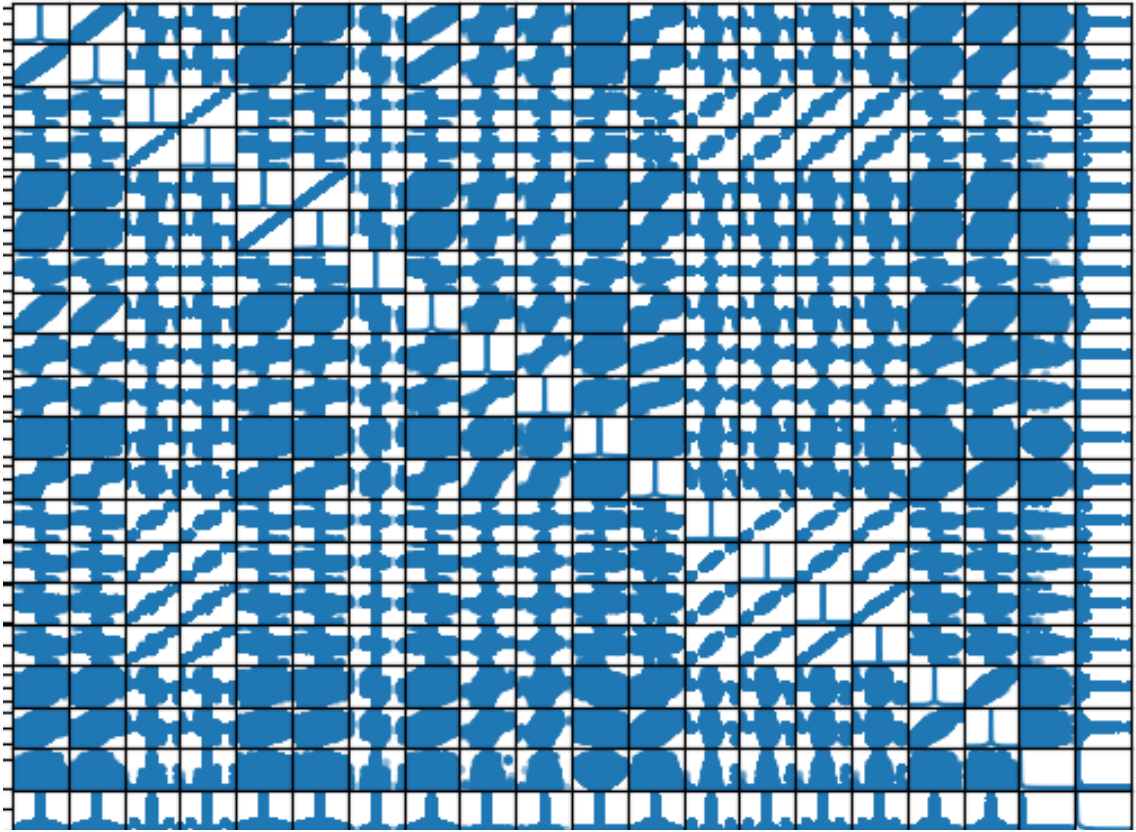

*Figure 8: Matrix of correlation and histograms for the gradient on network descriptors along user displacements. Descriptors in oder are: 'd_in_degree', 'd_out_degree', 'd_in_eigenvalue', 'd_out_eigenvalue', 'd_in_betweenness', 'd_out_betweenness', 'd_dis_betweenness', 'd_cfbetweenness', 'd_in_closeness', 'd_out_closeness', 'd_dis_closeness', 'd_cfcloseness0, 'd_in_ave_flow', 'd_out_ave_flow', 'd_in_std_flow' ,'d_out_std_flow', 'd_in_ave_distance', 'd_out_ave_distance', 'distance', 'flow'.*

The matrix of correlations and descriptor distribution is rather similar to the obtained from the relative displacements analysis, having the same correlation patterns (Fig. 8. The descriptors provide different information that is valuable to characterize the mobility.

Distribution of the statistics of the home-referenced vectors sequence also show a Guassian-like structure (Fig. 9). Interestingly, descriptors such as cfbetweeness show a different distribution between Bogotá and Medellín that was not detected in the analysis of relative displacements. This means that some descriptors on the home-referenced analysis may provide insights about the structure of the city and the distribution of where people live (Fig. 10)

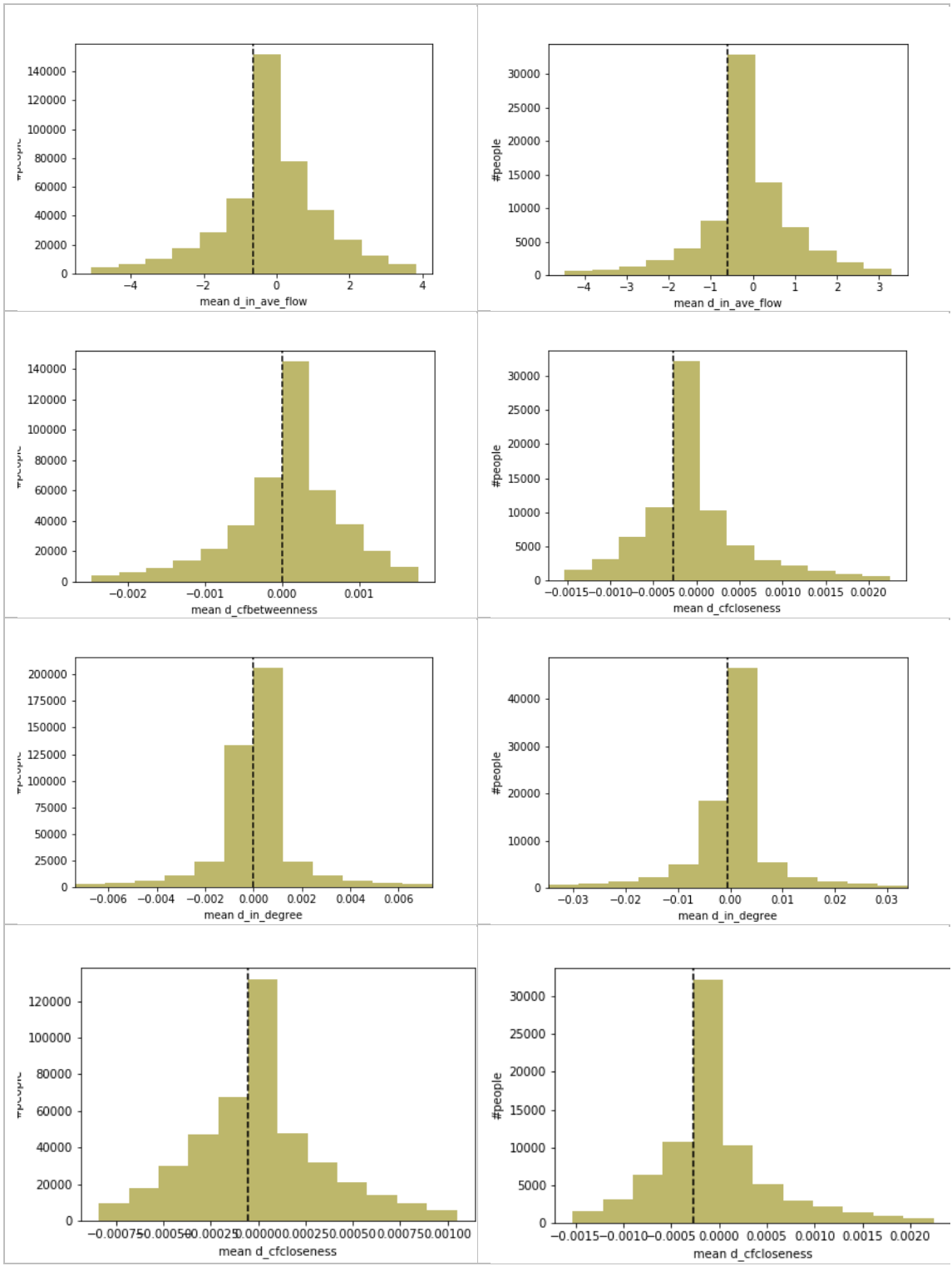

Figure 9: Histograms of the vectors home-referenced of topology descriptors. Right: Bogotá. Left: Medellín. First row: d_in_ave_flow. Second row: d_cfbetweenness. Third row: d_in_degree. Fourth row: d_cfcloseness

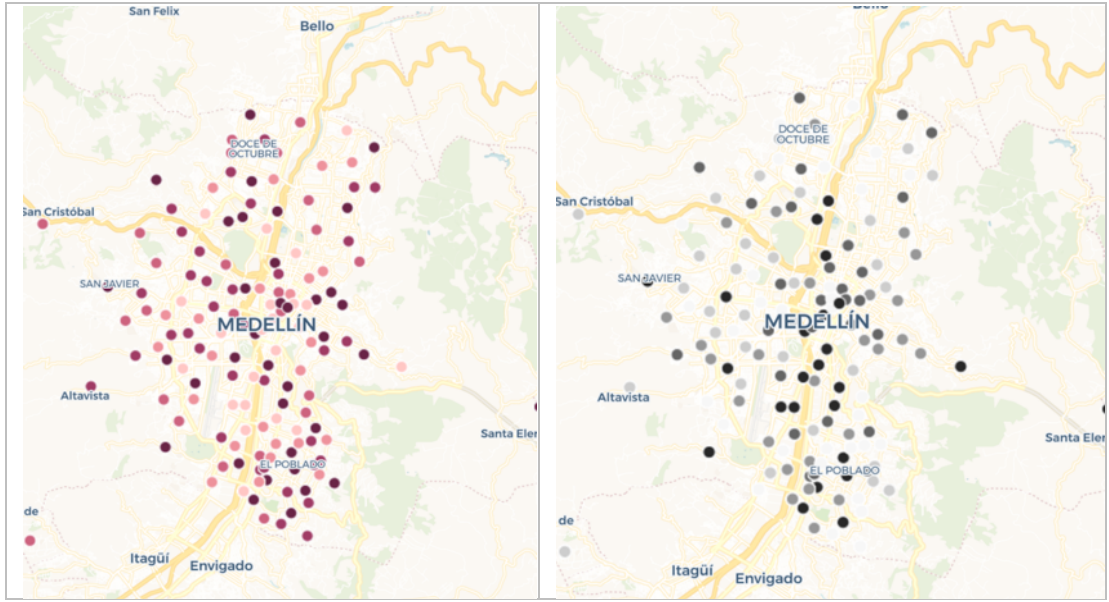

*Figure 10: Medellín. Left: average of cf_betweenness mean along users trajectories referred to the home location: Right: std of cf_betweenness mean along users trajectories referred to the home location*

**DISCUSSION**

The characterization of the network was based on the flows of people computed from displacements between geographical locations represented as network nodes. The flows were considered to be homogeneous (all people are considered equally to compute the flow), but heterogeneous flows could be also studied by labelling different types of populations leading to graphs with multivariate flow weights. For instance, by labelling population groups according to their cause of mobility or socio-economic or political status. Furthermore, the network nodes can be characterized in terms of communication flows.

The framework presented is suitable to understand the city layout and the mobility flows across areas of the city. It is also useful to make the assessment of variability and homophily within specific locations helping identify populations groups of interest for further analysis and mobility tracking. The framework can be also used to make a proper characterization of a baseline to measure the impact of crisis and disasters by monitoring large scale effects in the mobility network and how it may affect individuals. This framework is also relevant for more sophisticated modelling of human mobility that could be compared to gravity and radiation models for mobility (Simini et al., 2012) (Balcan et al., 2009) (Kang et al., 2015).

Further analysis involves creating a multi-scale spatial and temporal representation to better fit to the characteristics of the specific cities of study. City centres are often a nucleus of high flows of radial mobility masking other types of mobility within and between neighbourhoods with a lower flow magnitude. In this sense, a multi-scale approach of the flows magnitude could be also suitable to identify layers of mobility.

Nodes could classified according to contextual data and weight the importance of descriptor values. Thus, each node would be characterized with categorical and numerical variables. For instance, satellite data, business location data or demographics data would characterize the geographical location surrounding each antenna location to be combined with the topological role of the location.

This document sets the basis of a framework to investigate mobility flows integrating different scales and the effect of the network dynamics into individual mobility. Preliminary results show consistent insights and descriptive capabilities to characterize human mobility.

# SUPPLEMENTARY MATERIAL

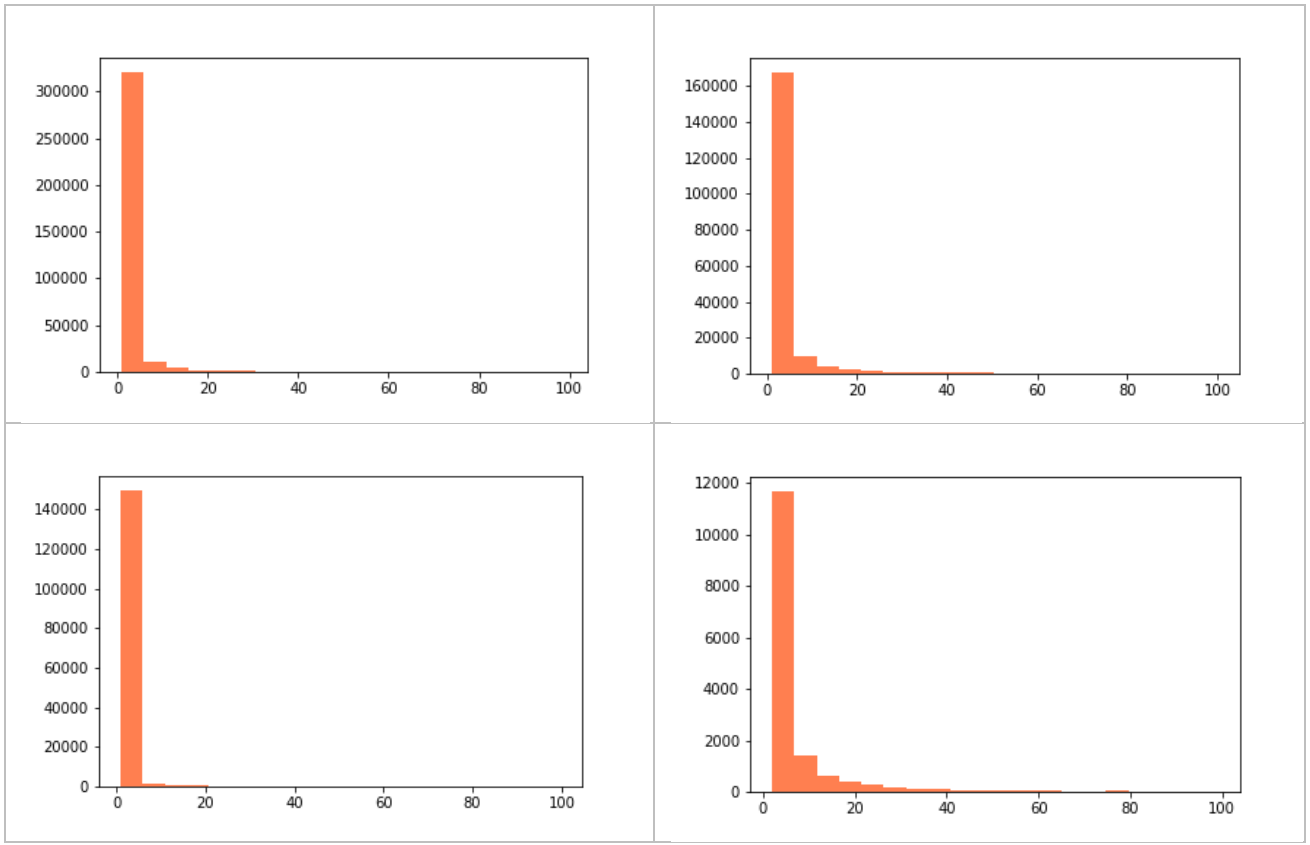

**Supplementary Figure 1: Distribution of network flows**

Top left: Histogram of flow magnitude in the network edges in the region of Bogotá.
Top right: Histogram of flow magnitude in the network edges in the region of Bogotá after removing all edges with a flow magnitude smaller than 2 people.
Bottom left: Histogram of flow magnitude in the network edges in the region of Medellín. Bottom right: Histogram of flow magnitude in the network edges in the region of Medellín after removing all edges with a flow magnitude smaller than 2 people.

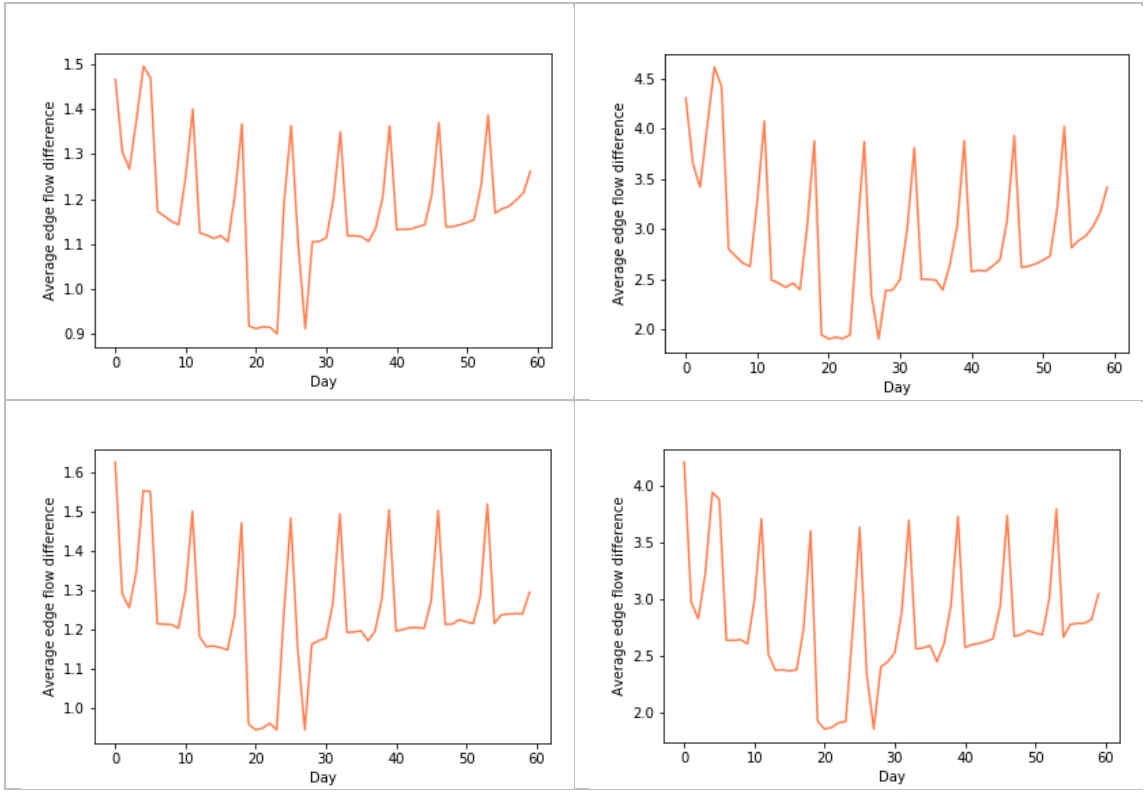

**Supplementary Figure 2: Comparison of reference network with daily network series**

Top left: Averaged flow difference between the edges of the reference network (unfiltered network) and the time series of the daily network in Bogotá. Top right: Averaged flow difference between the edges of the reference network (filtered network edges at least weight>=2) and the time series of the daily network in Bogotá. Bottom left: Averaged flow difference between the edges of the reference network (unfiltered network) and the time series of the daily network in Medellín. Bottom right: Averaged flow difference between the edges of the reference network (filtered network edges at least weight>=2) and the time series of the daily network in Medellín. The comparison temporal interval is two months: January and February 2014.

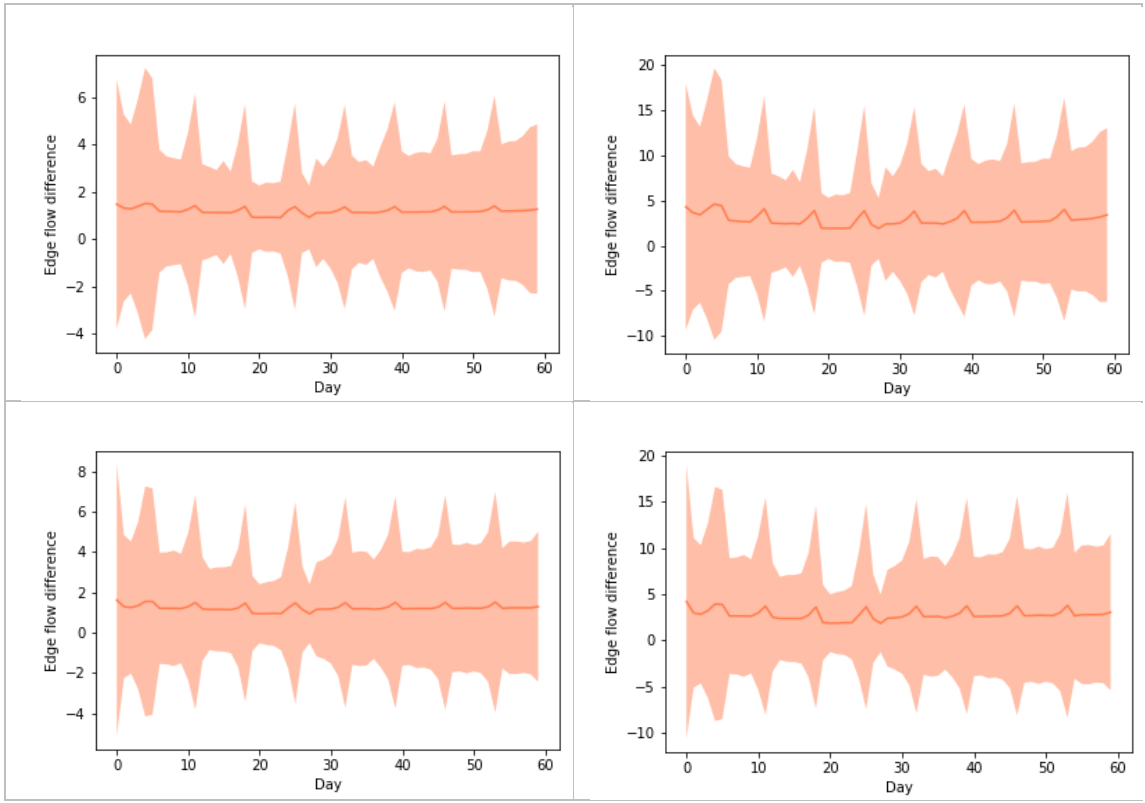

**Supplementary Figure 3: Comparison of reference network with daily network series**

Top left: Averaged flow difference and std between the edges of the reference network (unfiltered network) and the time series of the daily network in Bogotá. Top right: Averaged flow difference and std between the edges of the reference network (filtered network edges at least weight>=2) and the time series of the daily network in Bogotá. Bottom left: Averaged flow difference and std between the edges of the reference network (unfiltered network) and the time series of the daily network in Medellín. Bottom right: Averaged flow difference and std between the edges of the reference network (filtered network edges at least weight>=2) and the time series of the daily network in Medellín. The comparison temporal interval is two months: January and February 2014.

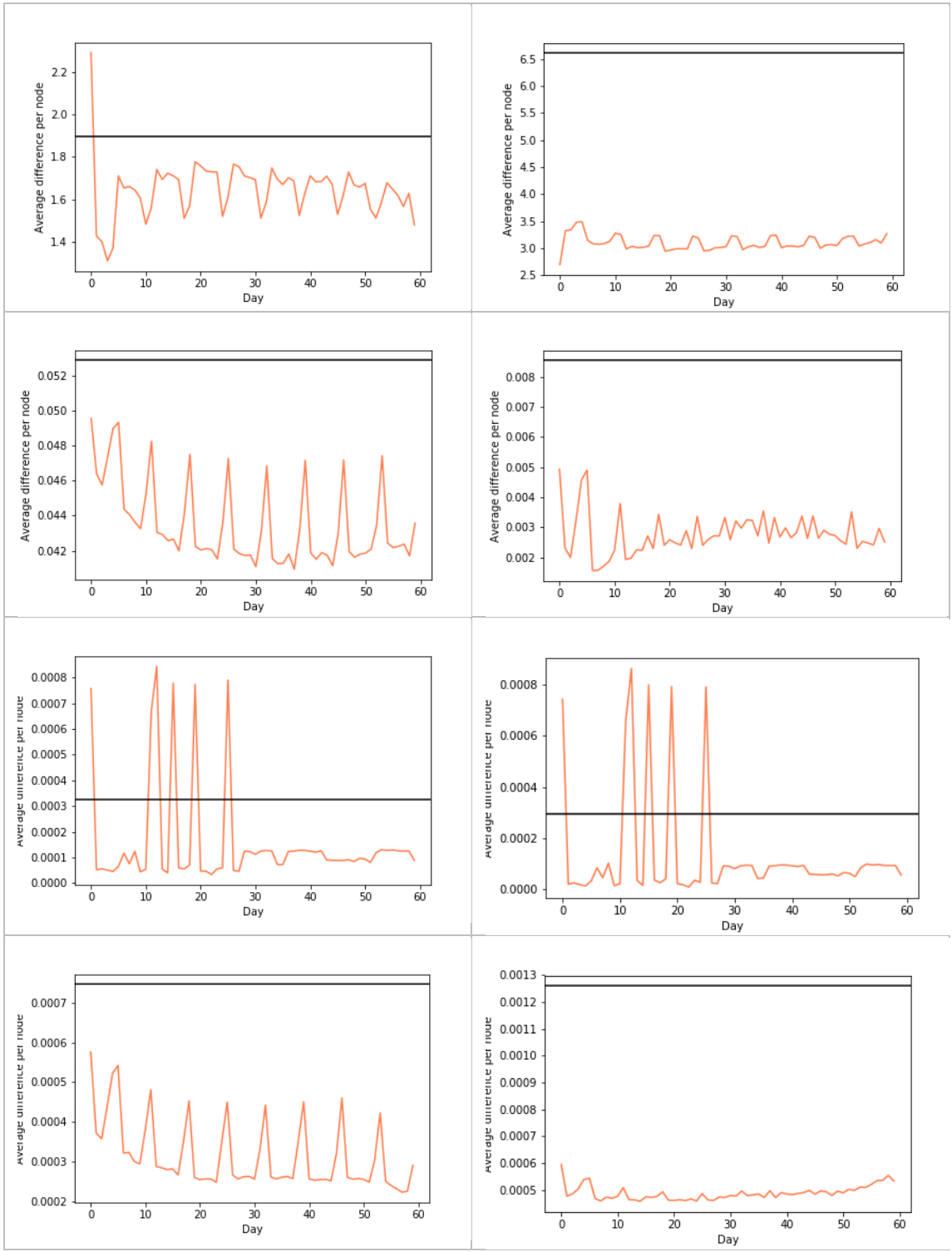

**Supplementary Figure 4: Comparison of reference network with daily network series**

First row: Averaged difference of out_ave_flow per node of the reference network (left unfiltered, right weight >= 2) and the time series of the daily out_ave_flow per node in Bogotá. Second row: Averaged difference of out_degree per node of the reference network (left unfiltered, right weight >= 2) and the time series of the daily out_degree per node in Bogotá. Third row: Averaged difference of out_eigenvalue per node of the reference network (left unfiltered, right weight >= 2) and the time series of the daily out_eigenvalue per node in Bogotá. Fourth row: Averaged difference of out_cfbetweenness per node of the reference network (left unfiltered, right weight >= 2) and the time series of the daily out_cfbetweenness per node in Bogotá.

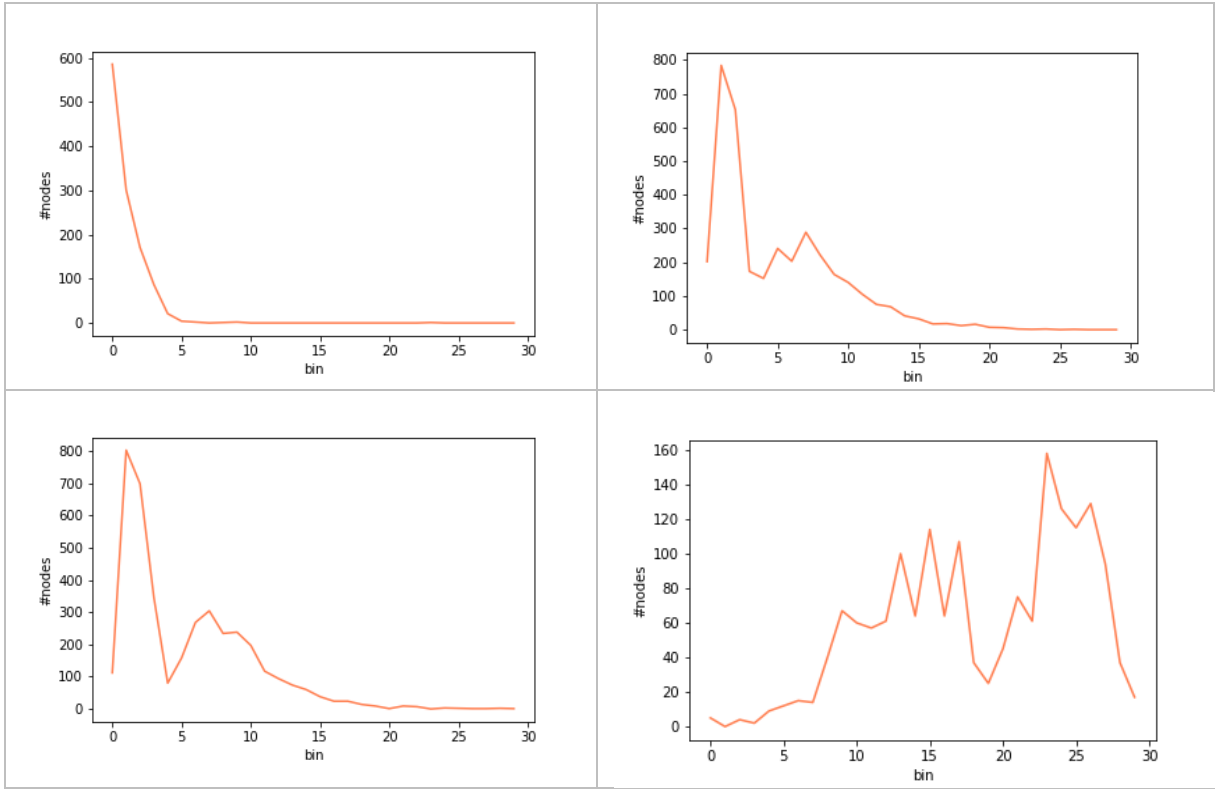

**Supplementary Figure 4: Comparison of reference networks: non filtered and filtered**

Histogram difference between centrality descriptors from the non-filtered network and the filtered network (weight >= 2). The histogram represents the distribution of descriptors for each node of the reference network. Top-left: in_ave_flow. Top-right: out_degree. Bottom-left: cfbetweenness. Bottom-right: cfcloseness.

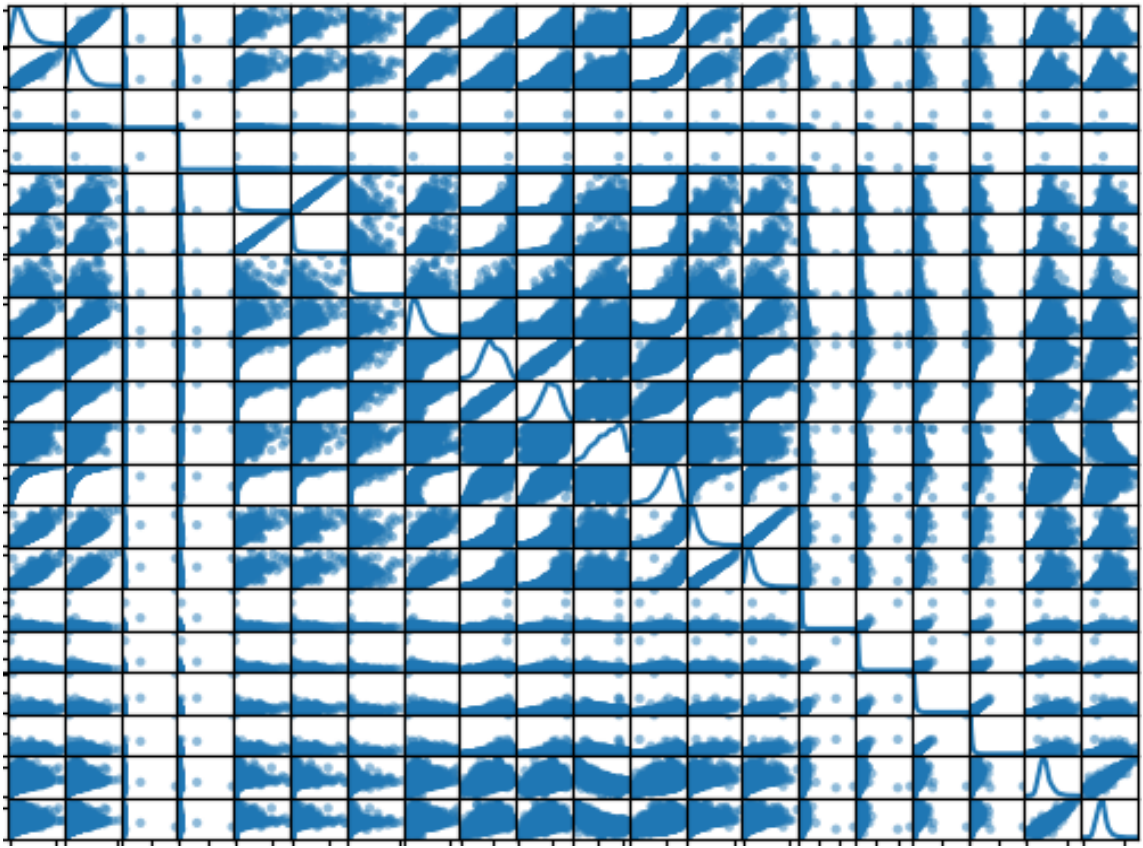

**Supplementary Figure 5: Descriptors characterization in the unfiltered reference network of Bogotá**

Exploration of descriptors histogram and pair-wise correlation at the antenna level in Bogotá. The twenty descriptors computed are: 'in_degree', 'out_degree', 'in_eigenvalue', 'out_eigenvalue', 'in_betweenness', 'out_betweenness', 'dis_betweenness', 'cfbetweenness', 'in_closeness', 'out_closeness', 'dis_closeness', 'cfcloseness', 'in_flow', 'out_flow', 'in_ave_flow', 'out_ave_flow', 'in_std_flow', 'out_std_flow', 'in_ave_distance', 'out_ave_distance'.

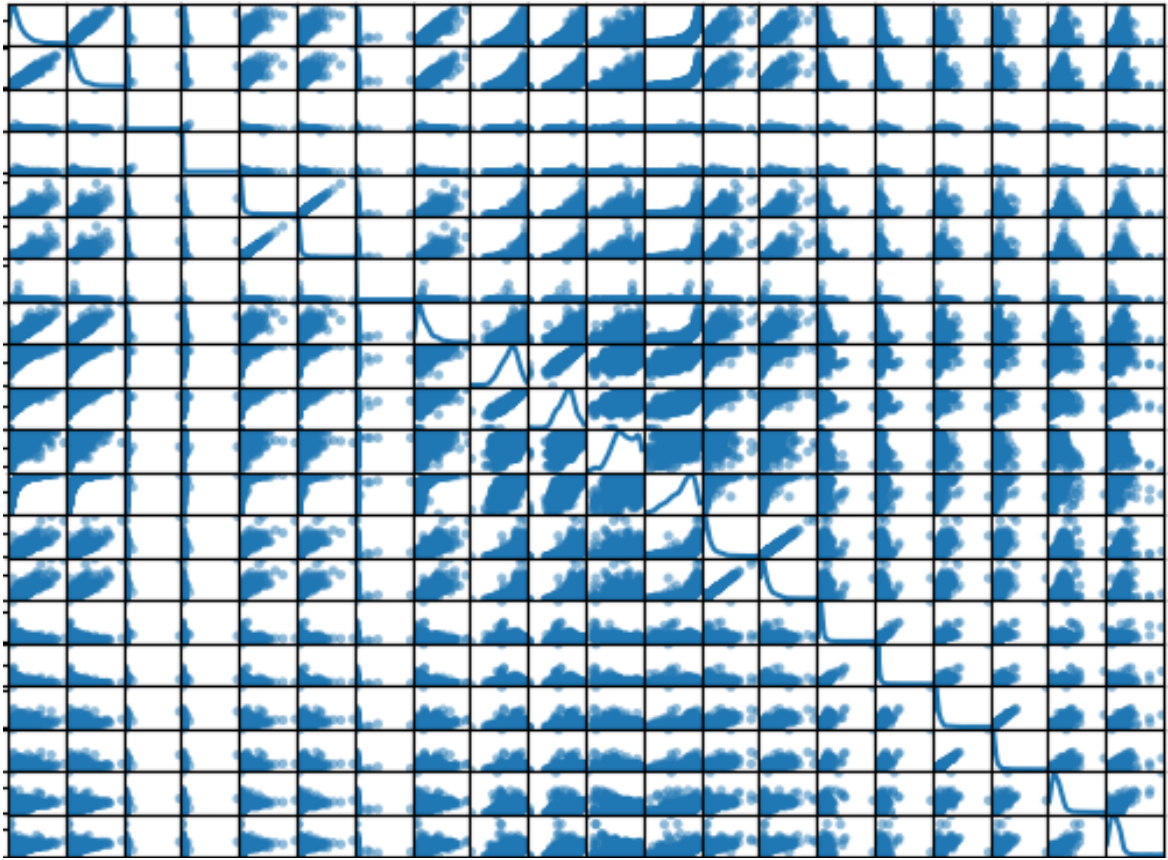

**Supplementary Figure 6: Descriptors characterization in the filtered reference network of Medellín**

Exploration of descriptors histogram and pair-wise correlation at the antenna level in Bogotá. The twenty descriptors computed are: 'in_degree', 'out_degree', 'in_eigenvalue', 'out_eigenvalue', 'in_betweenness', 'out_betweenness', 'dis_betweenness', 'cfbetweenness', 'in_closeness', 'out_closeness', 'dis_closeness', 'cfcloseness', 'in_flow', 'out_flow', 'in_ave_flow', 'out_ave_flow', 'in_std_flow', 'out_std_flow', 'in_ave_distance', 'out_ave_distance'.

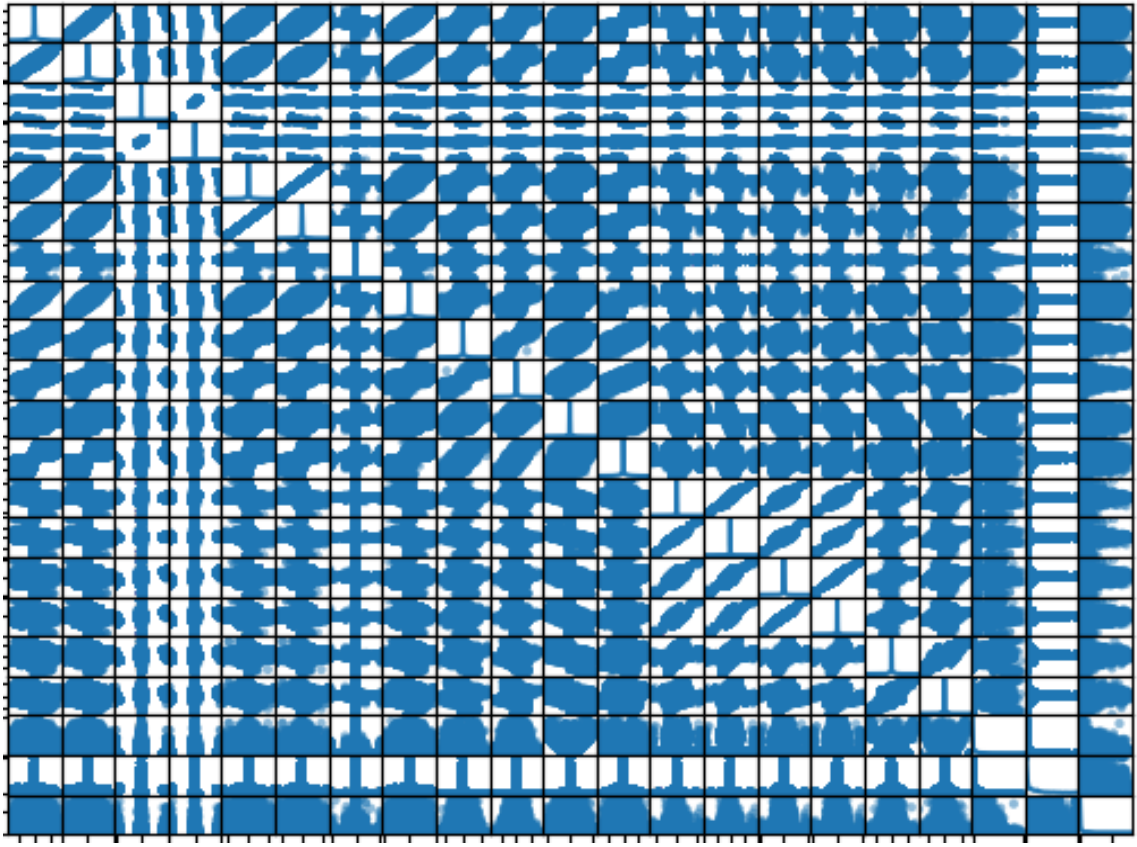

**Supplementary Figure 7: Matrix of correlation and histograms of the relative displacements vectorization in Medellín**

Matrix of correlation and histograms for the gradient on network descriptors along user displacements. Descriptors in oder are: 'd_in_degree', 'd_out_degree', 'd_in_eigenvalue', 'd_out_eigenvalue', 'd_in_betweenness', 'd_out_betweenness', 'd_dis_betweenness', 'd_cfbetweenness', 'd_in_closeness', 'd_out_closeness', 'd_dis_closeness', 'd_cfcloseness0, 'd_in_ave_flow', 'd_out_ave_flow', 'd_in_std_flow' ,'d_out_std_flow', 'd_in_ave_distance', 'd_out_ave_distance', 'distance', 'flow', 'd_time'.

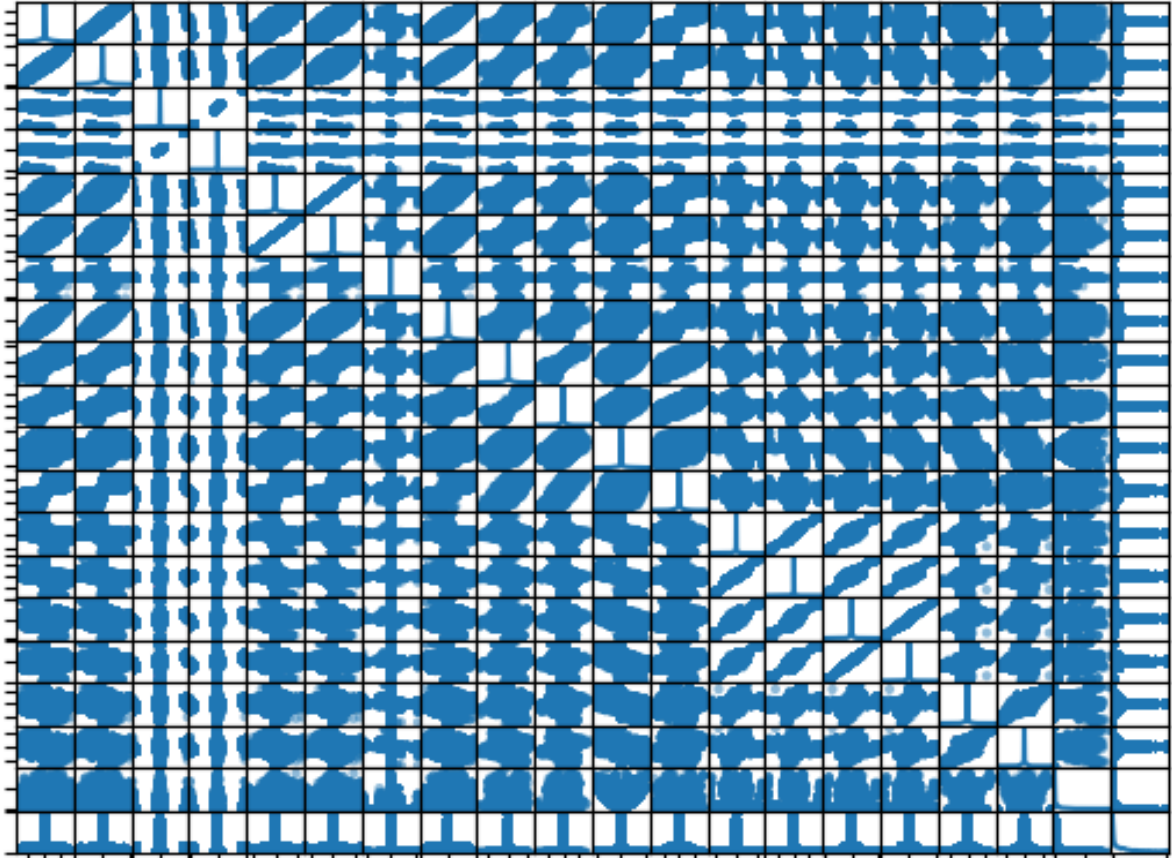

**Supplementary Figure 8: Matrix of correlation and histograms of the home-referenced vectorization in Medellín**

Matrix of correlation and histograms for the gradient on network descriptors along user displacements. Descriptors in oder are: 'd_in_degree', 'd_out_degree', 'd_in_eigenvalue', 'd_out_eigenvalue', 'd_in_betweenness', 'd_out_betweenness', 'd_dis_betweenness', 'd_cfbetweenness', 'd_in_closeness', 'd_out_closeness', 'd_dis_closeness', 'd_cfcloseness0, 'd_in_ave_flow', 'd_out_ave_flow', 'd_in_std_flow' ,'d_out_std_flow', 'd_in_ave_distance', 'd_out_ave_distance', 'distance', 'flow'.